\documentclass[a4paper,12pt]{article}

\usepackage{jheppub}

\usepackage[backgroundcolor=green!5, linecolor=blue!60]{todonotes}

\usepackage[section,above,below]{placeins}

\definecolor{lapis}{rgb}{0.0.0470,0.2941,0.5568}
\newcommand{\eq}[1]{\vspace{-0.5pt}\begin{equation}#1\vspace{-0.5pt}\end{equation}}

\usepackage{tikz}
\usetikzlibrary{arrows.meta}
\usetikzlibrary{fit}
\tikzset{%
  >={Latex[width=2mm,length=2mm]},
            base/.style = {rectangle, rounded corners, draw=black,
                           minimum width=1cm, minimum height=1cm,
                           text centered, font=\ttfamily},
            }

\usepackage[compat=1.1.0]{tikz-feynman}

\usepackage{xspace}

\newcommand{\Kira}{\texttt{Kira}\xspace}
\newcommand{\FIRE}{\texttt{FIRE}\xspace}
\newcommand{\DEAP}{\texttt{DEAP}\xspace}
\newcommand{\Python}{\texttt{Python}\xspace}
\newcommand{\funsearch}{\texttt{funsearch}\xspace}

\allowdisplaybreaks

\begin{document}

\title{Refining Integration-by-Parts Reduction of Feynman Integrals with Machine Learning}

\author[a]{Matt von Hippel,}
\affiliation[a]{Niels Bohr International Academy, Niels Bohr Institute, University of Copenhagen, \\ Blegdamsvej 17, 2100 Copenhagen \O{}, Denmark}
\author[a,b]{Matthias Wilhelm}
\affiliation[b]{%
Center for Quantum Mathematics, Department of Mathematics and Computer Science, University of Southern Denmark, Campusvej 55, 5230 Odense M, Denmark}

\date{\today}

\abstract{%
Integration-by-parts reductions of Feynman integrals pose a frequent bottle-neck in state-of-the-art calculations in theoretical particle and gravitational-wave physics, and rely on heuristic approaches for selecting integration-by-parts identities, whose quality heavily influences the performance.
In this paper, we investigate the use of machine-learning techniques to find improved heuristics. 
We use \funsearch, a genetic programming variant based on code generation by a Large Language Model, in order to explore possible approaches, then use 
strongly typed genetic programming 
 to zero in on useful solutions. Both approaches manage to re-discover the state-of-the-art heuristics recently incorporated into integration-by-parts solvers, and in one example find a small advance on this state of the art.
}

\maketitle

\section{Introduction}
\label{sec:introduction}

Perturbative Quantum Field Theory has proven to be a vastly successful theoretical framework for calculating precision predictions, with applications ranging from collider physics to gravitational-wave physics. A crucial step in the calculation of precision predictions is the reduction of the occurring Feynman integrals to a much smaller set of so-called master integrals, using integration-by-parts (IBP) identities~\cite{Tkachov:1981wb,Chetyrkin:1981qh,Laporta:2000dsw}.
This IBP reduction is a major bottleneck in precision calculations, requiring hundred thousands of CPU hours in current applications~\cite{Driesse:2024xad} and obstructing other applications altogether. 

IBP identities relate Feynman integrals with different integer exponents of the propagators as well as irreducible scalar products (ISP) in the numerator. They can easily be derived for general values of the exponents, see e.g.\ ref.~\cite{Weinzierl:2022eaz} for a textbook treatment. In contrast, it is in most cases not possible to solve the resulting systems of IBP identities in closed form, i.e.\ for general values of the exponents. Instead, IBP reduction codes such as \texttt{AIR} \cite{Anastasiou:2004vj}, \FIRE \cite{Smirnov:2008iw,Smirnov:2023yhb}, \texttt{Reduze} \cite{vonManteuffel:2012np}, \texttt{LiteRed} \cite{Lee:2012cn}, \Kira \cite{Maierhofer:2017gsa,Klappert:2020nbg}, \texttt{FiniteFlow} \cite{Peraro:2019svx} and \texttt{Blade} \cite{Guan:2024byi} specialize the identities to a sufficiently large set of different values for the integer exponents -- the so-called seeds -- to solve for the desired integrals in terms of the master integrals. 
The choice of the seeds is determined by a so-called seeding strategy, a heuristic whose quality heavily influences the performance of IBP reduction.
Recently, a new heuristic was proposed~\cite{JohannQCDmeetsGravity,Driesse:2024xad,Guan:2024byi,Bern:2024adl}, which reduces the size of the resulting system of linear equations and thus the reduction time by orders of magnitude. 
However, an optimal choice of seeds is in general not known.%
\footnote{There exist attempts to improve IBP reduction by choosing a particular suitable basis in the space of IBP identities via syzygy methods, see refs.~\cite{Gluza:2010ws,Wu:2023upw} and references therein, which are currently being implemented in IBP codes. Other approaches aim to replace IBP reduction via intersection theory \cite{Mastrolia:2018uzb,Frellesvig:2019uqt} but have not produced competitive reduction codes yet.
}

Motivated by the discovery of the improved heuristic in refs.~\cite{JohannQCDmeetsGravity,Driesse:2024xad,Guan:2024byi,Bern:2024adl},
in this paper we employ automated, machine-learning based methods to search for further improved heuristics and optimal seeding strategies. Concretely, we use three different versions of genetic algorithms, two of which involve genetic programming.

Genetic algorithms imitate evolution by natural selection; see e.g.\ ref.~\cite{evolutionary_programming} for a textbook treatment. They involve a population of candidate solutions to a problem. This population is subject to mutation, in which individuals are randomly altered, and cross-over, in which pairs of individuals give rise to new members of the population with traits drawn from each parent, imitating sexual reproduction. The population is then selected based on its performance by some evaluation metric, with the best individuals kept for the next generation.

In a subset of genetic algorithms, called genetic programming, the individuals in the population are programs, parametrized e.g.\ with trees; see e.g.\ ref.~\cite{genetic_programming}. A recent example of genetic programming, called \funsearch~\cite{funsearch}, instead parametrizes programs in terms of text, more specifically \Python code. Mutation and cross-over of texts are then carried out using a Large Language Model (LLM). The LLM is given the \Python code representing two individuals in the population, then asked how to improve on it. 
\funsearch has shown some success in finding novel solutions to problems in pure mathematics~\cite{funsearch}. It has the advantage that one typically needs to know very little about the problem one is applying it to, as solutions are written in \Python code, not in any specialized framework, making it well-suited to exploratory work.
Moreover, the output of \funsearch being \Python code makes it easily interpretable, and thus generalizable.

Throughout this paper, we use the two-loop triangle-box integral depicted in fig.~\ref{fig:triangle} as a benchmark to measure to performance of candidate seeding strategies. This integral was already used as a benchmark in ref.~\cite{JohannQCDmeetsGravity}.
We begin by trying a traditional genetic algorithm, treating the list of seed integrals used as a binary vector. This approach is very slow to converge. We then explore the problem using \funsearch. With some judicious prompting, \funsearch is able to find solutions that not only reach the state of the art, but in an example find slight improvements on it. We then use the insights we gained via \funsearch to proposed a more specialized list of operations appropriate to our problem, running these via strongly type genetic programming as implemented in the genetic algorithm library \DEAP~\cite{deap}. Strongly typed genetic programming converges much faster than \funsearch, finding the same best solution in only thirty generations.

The remainder of this paper is structured as follows. In the next section, we will provide background on integration-by-parts methods for Feynman integrals in subsection~\ref{subsec:ibp} and background on genetic algorithms, genetic programming, and machine learning in subsection~\ref{subsec:gaml}. We then describe our initial attempt with a genetic algorithm operating on binary vectors in section~\ref{sec:ga}, before describing our use of \funsearch in section~\ref{sec:funsearch} and the strongly typed genetic programming approach in section~\ref{sec:gp}. Finally we conclude and discuss ways in which these approaches might be used in the future in section \ref{sec:conclusion}.
We provide the identities used for the reduction of our benchmark Feynman integral in appendix~\ref{sec:fullrelations}. 

\section{Background}
\label{sec:background}

In this section, we provide a brief introduction to integration-by-part identities for Feynman integrals as well as to genetic algorithms. 

\subsection{Integration-by-Parts identities for Feynman Integrals}
\label{subsec:ibp}

In perturbative Quantum Field Theory beyond the leading order, one typically faces the task of evaluating a large number of Feynman integrals. 
We give a brief introductions to these integrals here, illustrated with one example; see e.g.\ ref.~\cite{Weinzierl:2022eaz} for a detailed text-book treatment. 

For each Feynman integral, we have an associated graph with $E$ external edges as well as further internal edges.
Associated to each external edge is an external momentum vector $p_j^\mu$ with $j=1,\dots,E$ and $\mu=0,\dots,D-1$, a vector in $D$-dimensional Minkowski space. These vectors satisfy so-called momentum conservation, $\sum_{j=1}^Ep_j^\mu=0$.  
The integration in a Feynman integral is with respect to $L$ loop momentum vectors $k_l^\mu$ ($l=1,\dots,L$) in $D$-dimensional Minkowski space, corresponding to the $L$ cycles in the associated graph.
We organizing Feynman integrals into families indexed by sets of integers $a_1, \dots, a_n$, where $n=L(L+1)/2+L(E-1)$ is the number of independent loop-momentum-dependent products that can be formed from the momenta. The general form of an integral family is 
\eq{
I(a_1,\ldots ,a_n)=\int\frac{\prod_{l=1}^L d^D k_l}{\prod_{i=1}^n [D_i(k_1^\mu,\ldots,k_L^\mu)]^{a_i}}\,,
}
where the polynomials $D_i\equiv D_i(k_1^\mu,\ldots,k_L^\mu)$ are always non-trivial functions of at least one of the loop momenta $k_l^\mu$, and can also be functions of the external momenta and masses. For the vast majority of applications, the $D_i$ are quadratic in these variables and Lorentz invariant. The dimension of space-time $D$ is usually taken to be non-integer in a regularization technique called dimensional regularization \cite{tHooft:1972tcz}. 
Note that the $a_i$, used to index a family with a common set of $D_i$, can be positive, in which case the corresponding $D_i$ is a propagator associated to an edge of the graph. Indices $a_i$ that are larger than one correspond to higher powers of the propagator.
Moreover, $a_i$ can be negative, corresponding to irreducible scalar products (ISPs) in the numerator, or zero, in which case $D_i$ is absent in that member of the family.

\begin{figure}[t]
\begin{center}
    \begin{tikzpicture}
    \begin{feynman}
      \vertex (a) at (-4,0);
      \vertex (b) at ( -2,0);
      \vertex (c1) at (0,1);
      \vertex (c2) at (0,-1);
      \vertex (d1) at (2,2);
      \vertex (d2) at (2,-2);
      \vertex (e1) at (4,3);
      \vertex (e2) at (4,-3);
        \diagram* {
    (a) -- [very thick,rmomentum=$p_1$] (b),
    (b) -- [very thick,momentum=$k_1$] (c1),
    (b) -- [very thick,rmomentum'=$k_1+p_1$] (c2),
    (c1) -- [very thick,momentum=$k_1+k_2$] (c2),
    (c1) -- [very thick,rmomentum=$k_2$] (d1),
    (c2) -- [very thick,momentum'=$k_2-p_1$] (d2),
    (d1) -- [very thick,rmomentum=$k_2+p_3$] (d2),
    (d1) -- [very thick,momentum=$p_3$] (e1),
    (d2) -- [very thick,momentum'=$p_2$] (e2),
   };
    \end{feynman}
    \end{tikzpicture}
\end{center}
\caption{The two-loop triangle-box integral we used to benchmark different heuristics. 
The quantities at the arrows specify the momenta flowing through the edges of the graph. 
}
\label{fig:triangle}
\end{figure}
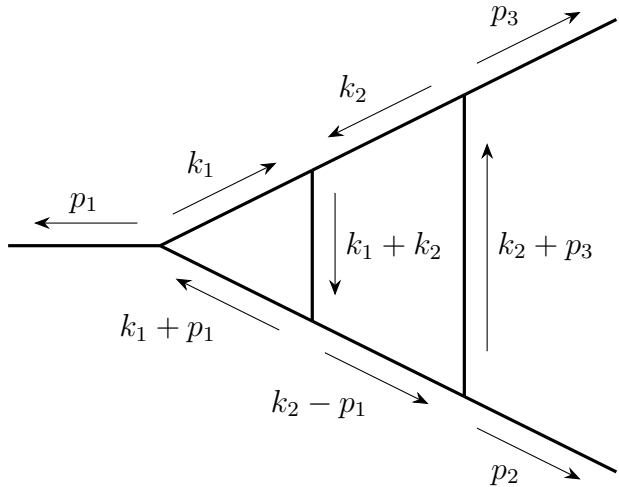
To illustrate, consider the Feynman integral depicted in fig.~\ref{fig:triangle}. This integral was one of the examples used to demonstrate the advantages of the improved seeding strategy in ref.~\cite{JohannQCDmeetsGravity}, and we will use it to benchmark our machine-learning approaches throughout this paper.
It has two loops, $L=2$, and three external momenta, $E=3$, and thus $n=7$.
We can parametrize the seven $D_i$ as follows:
\begin{align}
D_1 = k_1^2\,,\quad D_2= k_2^2\,,  \quad &D_3=(k_1+k_2)^2\,,\quad D_4=(k_1+p_1)^2\,,\nonumber\\
D_5=(k_2+p_3)^2\,,\quad D_6 &=(k_2-p_1)^2\,,\quad D_7=(k_1+p_3)^2\,.
\end{align}
The first six of these polynomials, $D_1,\ldots, D_6$, correspond to internal edges in the graph depicted in fig.~\ref{fig:triangle}, and are squares of the momenta flowing through those edges. For a Feynman integral depicted by the graph in fig.~\ref{fig:triangle}, these quantities will appear in the denominator of the corresponding integrand, so $a_1,\ldots, a_6$ will be positive. For the methods described in this section to work, the full set of $D_i$ must form a basis for quadratic polynomials in the $k_l^\mu$ and $p_j^\mu$, up to terms that can be expressed only in terms of Lorentz-invariant combinations of external momenta. As $D_1,\ldots, D_6$ do not constitute such a basis on their own, we must add $D_7$, which is thus referred to as an ISP, and is an example of the ISPs mentioned above. These will not occur as denominators in this family of integrals but may appear as numerators, so the corresponding $a_7$ will be zero or negative. 

The Feynman integral depicted in fig.~\ref{fig:triangle} depends on the invariant quantities $p_1^2=m_1^2$, $p_2^2=m_2^2$ and $p_3^2=m_3^2$ that can be formed from the three external momenta $p_1^\mu$, $p_2^\mu$ and $p_3^\mu$ with $p_1^\mu+p_2^\mu+p_3^\mu=0$. Here, the square $p_i^2$ denotes the (pseudo-) norm with respect to the Lorentz product. The overall dependence on these dimensionful quantities is determined by dimensional analysis, such that the integral has a non-trivial dependence only on the two dimensionless ratios $m_2/m_1$ and $m_3/m_1$. For simplicity, we will thus set $m_1\equiv 1$.

Within dimensional regularization, any integral that is independent of both all non-zero Lorentz invariants composed from the external momenta and all of the masses is set to zero, and this condition can be enforced independently for the integration over each loop momentum; see ref.~\cite{Abreu:2022mfk} for a pedagogical review that should clarify some topics discussed here for mathematical readers. These integrals are called ``trivial''. 

Not all integrals built from a given set of $D_i$ are linearly independent. They are related by integration-by-parts (IBP) identities \cite{Tkachov:1981wb,Chetyrkin:1981qh} generated by expressions of the form%
\footnote{Here, we use Einstein's summation convention, i.e.\ the sum over the repeated index $\mu$ is implied.}
\eq{
0=\int \prod_{i=1}^L d^D k_i \frac{d}{dk_l^\mu}\frac{q^\mu}{\prod_{i=1}^n D_i^{a_i}}\,,
\label{eq:basic_ibp}}
where the so-called IBP vector $q^\mu$ is built from external momenta and loop momenta.
While eq.~\eqref{eq:basic_ibp} holds for any choice of IBP vector, it is sufficient to consider $q^\mu\in\{p_j^\mu,k_l^\mu\}$ to obtain a generating set of all IBP identities, and we will do so in this paper. Note, however, that more tailored choices of IBP vectors can be advantageous, as used by syzygy methods~\cite{Gluza:2010ws,Wu:2023upw}. 
Moreover, there are Lorentz invariance (LI) identities generated by acting on Feynman integrals with generators of Lorentz transformations. These identities are not independent from the IBP identities~\cite{Lee:2008tj}, but are often useful to include, and we will include them in the examples in this paper.

Provided one chooses the set $D_i$ so as to be a basis for ISPs of the loop momenta with each other and the external momenta, applying the product and chain rules in eq.~\eqref{eq:basic_ibp} generates linear relations between Feynman integrals $I(a_1,\ldots,a_n)$ with shifted exponents $a_i$, with coefficients that are rational functions in the masses, the dimension, and ISPs of the external momenta (together called kinematic parameters).
One wants to solve these relations, so as to represent all of the integrals that appear in a given calculation in terms of a minimal basis of so-called master integrals.
 
 In our example, there are eight IBP identities -- two loop momenta $k_l$ times four total independent momenta $\{k_1,k_2,p_1,p_2\}$ that can serve as IBP vectors -- and one identity generated by Lorentz transformations. An example of an IBP identity is
 \begin{align}
&(D{-}a_2{-}a_6{-}a_3{-}2a_0)I(a_1,a_2,a_3,a_4,a_5,a_6,a_7)-
a_2 I(a_1-1,a_2,a_3+1,a_4,a_5,a_6,a_7)\nonumber\\*
&\quad +
a_2 I(a_1,a_2-1,a_3+1,a_4,a_5,a_6,a_7) +
a_6 m_3^2 I(a_1,a_2,a_3,a_4,a_5,a_6,a_7+1)\nonumber\\*
&\quad -
a_3 I(a_1-1,a_2,a_3,a_4+1,a_5,a_6,a_7) -
a_6 I(a_1-1,a_2,a_3,a_4,a_5,a_6,a_7+1)\nonumber\\*
&\quad +
a_3 I(a_1,a_2,a_3,a_4+1,a_5,a_6,a_7)=0\,.
 \end{align}
 We give the full set of identities for this integral in appendix~\ref{sec:fullrelations}.
 When these identities are solved, all integrals in this family can be expressed in terms of a basis of 16 master integrals. An example choice of such basis is 
 \begin{align}
 &I(0,1,1,1,0,0,0), \, 
&&I(1,0,1,0,1,0,0),  \,
&&I(1,1,0,1,1,0,0),  \,
&&I(0,0,1,1,1,0,0),  \nonumber\\
&I(1,0,1,1,1,0,0), \, 
&&I(1,-1,1,1,1,0,0), \,
&&I(0,1,1,1,1,0,0),  \,
&&I(-1,1,1,1,1,0,0), \nonumber\\
&I(1,1,1,1,1,0,0), \, 
&&I(1,0,1,0,0,1,0),  \,
&&I(1,1,0,1,0,1,0),  \,
&&I(1,0,1,0,1,1,0), \nonumber\\ 
&I(1,-1,1,0,1,1,0),\, 
&&I(1,0,0,1,1,1,0),  \,
&&I(1,1,0,1,1,1,0),  \,
&&I(1,0,1,1,1,1,0) \,.
\label{eq: master integrals example}
 \end{align}

Sometimes, it is possible to solve the systems of IBP identities in full generality for all values of $a_i$; see ref.~\cite{Lee:2012cn} for heuristic code to find these solutions.\footnote{See also ref.~\cite{Kosower:2018obg}.} However, there is no known algorithm that can find these solutions in all cases. As such, in practice one often instead solves a system generated by a finite list of seeds $a_1,\ldots,a_n$ in eq.~\eqref{eq:basic_ibp}.
If this finite list is large enough, it will still allow the reduction of the integrals of interest in terms of a minimal basis. 
In practice, seeds corresponding to trivial sectors are often excluded, and we also do this here. 

For later convenience, we define the following quantities for a given member of an integral family,  $I(a_1,\ldots,a_n)$:
\begin{align}
t\equiv \sum_{a_i>0} 1 \,,&&
r\equiv \sum_{a_i>0} a_i \,,&&
d\equiv r-t = \sum_{a_i>0} (a_i-1)\,,&&
s\equiv-\sum_{a_i<0} a_i \,.
\end{align}
The quantity $t$ counts the total number of propagators, $r$ the sum of propagator powers, $d$ the total number of propagator repetitions, and $s$ the total numerator power. Since repetitions of propagators are typically depicted by dots on the respective edge in the graph, $d$ is also referred to as the total number of dots.
In any given problem one typically has one so-called top sector in each integral family, giving a maximal list of which $a_i$ are allowed to be positive; the other $a_i$ will always be zero or negative. Thus there is a maximum value for $t$, $t_{\textrm{max}}$, determined by the top sector.

A number of heuristic seeding strategies have been proposed by different authors and implemented in different IBP codes, starting with Laporta's \emph{golden rule}~\cite{Laporta:2000dsw}. We will refer to these strategies as follows:%
\footnote{While Laporta's \emph{golden rule} was proposed first~\cite{Laporta:2000dsw}, more recent implementations of IBP codes have been using rectangular seeding \cite{Maierhofer:2018gpa} before the improved seeding strategy was discovered \cite{JohannQCDmeetsGravity,Driesse:2024xad,Guan:2024byi,Bern:2024adl}.
} 
\begin{itemize}
\item \textit{Rectangular Seeding:} Use all seeds $a_1,\ldots,a_n$ such that $r\leq r_{\textrm{max}}$ and $s\leq s_{\textrm{max}}$ for choices of $r_{\textrm{max}}$ and $s_{\textrm{max}}$ that include the integrals of interest.
\item \textit{Golden Rule:} Use seeds constrained as above, and also demand that $d\leq d_{\textrm{max}}$ for a choice of $d_{\textrm{max}}$ that includes the integrals of interest, so that the integrals corresponding to the seeds keep the same number of propagator repetitions in lower sectors. 
\item \textit{Improved Seeding:} Use seeds constrained as above, and also require $s\leq t-l+1$ for a choice of parameter $l$ that includes the integrals of interest, so that integrals in lower sectors also have fewer powers of ISPs in the numerator.
\end{itemize}
For future reference, we will mention that in the case that $d_\textrm{max}=0$, the additional condition imposed by improved seeding can be simply written as $\sum_i a_i \geq l-1 $.

In general one wants to define $r_\textrm{max}$, $s_{\textrm{max}}$, and $d_{\textrm{max}}$ to be as small as possible while still including the integrals of interest. However, Laporta observed certain minimal values for these parameters below which one does not achieve a complete reduction~ \cite{Laporta:2000dsw}. In particular, he noted that one sometimes needs to take at least $d_{\textrm{max}}=1$, even if $d=0$ for all integrals of interest.

In the example we use to benchmark different approaches throughout this paper, as in ref.~\cite{JohannQCDmeetsGravity}, we would like to reduce the integral $I(1,1,1,1,1,1,-3)$ to master integrals, so we need $s_\textrm{max}=3$. We consider cases with either $r_\textrm{max}=6$ or $r_\textrm{max}=7$, following Laporta's observation that one sometimes needs $d_{\textrm{max}}=1$ even if one is only interested in integrals with $d=0$.
To gain an intuition about the size of the corresponding systems of IBP equations, we list some examples:
\begin{itemize}
\item Rectangular seeding with $s_\textrm{max}=3$ and $r_\textrm{max}=7$ yields 14,588 seeds.
\item The golden rule with $s_\textrm{max}=3$, $r_\textrm{max}=7$, and $d_\textrm{max}=1$ yields 2,148 seeds.
\item Improved seeding with $s_\textrm{max}=3$, $r_\textrm{max}=6$, $d_\textrm{max}=0$, and $l=4$ yields 92 seeds.
\end{itemize}

As mentioned above, IBP systems for Feynman integrals are systems of linear equations with rational function coefficients. Historically, such systems were solved using symbolic algebra. However, this proved excessively cumbersome for larger systems, and in recent years the community has instead turned to methods using finite fields~\cite{vonManteuffel:2014ixa,Peraro:2016wsq,Klappert:2019emp,Peraro:2019svx}. 
By substituting in integers for the kinematic parameters, one can solve the IBP system over a finite (large prime) field with much less computational cost than solving the full symbolic system. If one does this for a sufficient number of different points, one can use finite field reconstruction techniques to determine the rational functions present in the solution. Meanwhile, solving at just a single point is enough to identify a list of master integrals for the system, provided the point is sufficiently generic.

Throughout this paper, once we have generated a set of seeds and the resulting system of IBP equations, we use \Kira~\cite{Maierhofer:2017gsa,Klappert:2020nbg} to check that the system that we have generated is sufficiently large to provide a full reduction of the target integral $I(1,1,1,1,1,1,-3)$ in terms of the master integrals.
To do this, we have \Kira perform only its initialization step, which uses \texttt{pyRed} to solve the system over a finite field at a single kinematic point, and check whether the number of master integrals is equal to the number $16$ found by solving the rectangular system.

\subsection{Genetic Algorithms and Machine Learning}
\label{subsec:gaml}

Genetic algorithms are heuristic methods used to search for solutions that score well on an evaluation metric by emulating evolution via natural selection. They begin with 
a population of individuals, represented by a DNA-like string of symbols (typically, integers), their ``genotype''. These genotypes are mutated, and crossed-over i.e.\ the genotypes of two individuals are mixed, to create a new population, from which the fittest elements (i.e.\ the ones with the best evaluation metric) are selected. This process is repeated over a number of generations, in the hope of creating populations with better evaluation metric, while still retaining the genetic diversity that allows for further improvement.
See e.g.\ ref.~\cite{evolutionary_programming} for a textbook introduction to genetic algorithms. 

Genetic algorithms have several features which have motivated a variety of methods. We mention several, highlighting the methods used in parts of this work:
\begin{itemize}
\item  There are many ways to select for the best individual, and selection can happen both on the parent (who will mate) and on the children (who will survive). Instead of strictly selecting the best individuals, one can select individuals randomly with probability weighted by their fitness in various ways. Examples include roulette selection, where probabilities are strictly weighted by their fitness, tournament selection, where individuals are compared in random smaller groups and only the winners from those groups are preserved, and Boltzmann selection, in which individuals are selected via a thermal partition function \cite{10.5555/645513.657604}.
\item  It is observed that population diversity tends to decrease from one generation to the next. As a cure, instead of just maintaining a single population, one can maintain a number of islands, sub-populations which are independently subject to crossover and selection. These islands interact with each other more rarely, for example by occasionally removing the worst-performing islands and re-populating them from the best-performing ones. Islands can help to preserve a greater diversity of individuals in order to explore a wider range of possibilities.
\item The method heavily depends on the representation chosen for the genotype, and the operators chosen for cross-over.
 For example, one could randomly choose elements of the child to come from one or the other parent, or one could form the child by splicing part of one parent with the complementary part of another, randomly choosing the position of the splice.
\end{itemize}

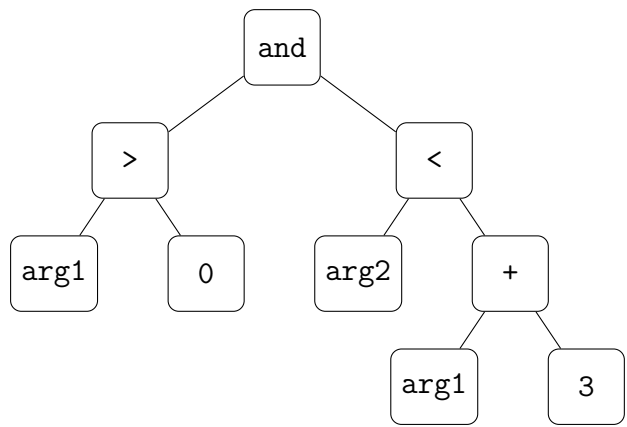
\begin{figure}[t]
\centering
\begin{tikzpicture}[node distance=1.5cm,
    every node/.style={fill=white, font=\sffamily}, align=center]

\node(and) [base] {and};

\node(greater) [base,below of=and, xshift=-2cm] {>};

\draw            (and) -- (greater);

\node(lesser) [base,below of=and, xshift=2cm] {<};

\draw            (and) -- (lesser);

\node(arg1) [base,below of=greater, xshift=-1cm] {arg1};

\draw            (greater) -- (arg1);

\node(zero) [base,below of=greater, xshift=1cm] {0};

\draw            (greater) -- (zero);

\node(arg2) [base,below of=lesser, xshift=-1cm] {arg2};

\draw            (lesser) -- (arg2);

\node(plus) [base,below of=lesser, xshift=1cm] {+};

\draw            (lesser) -- (plus);

\node(arg1p) [base,below of=plus, xshift=-1cm] {arg1};

\draw            (plus) -- (arg1p);

\node(three) [base,below of=plus, xshift=1cm] {3};

\draw            (plus) -- (three);

\end{tikzpicture}
\caption{A tree diagram for a simple program given in the main text.}
\label{fig:exampletree}
\end{figure}

Genetic programming is a particular use of genetic algorithms, in which the individuals in the population each represent a program; see e.g.\ ref.~\cite{genetic_programming} for an introduction. 
Typically, these programs are represented as trees of operations. For example, the \Python function
\begin{verbatim}
def func(arg1,arg2):
	return arg1>0 and arg2<arg1+3
\end{verbatim}
can be represented by the tree shown in fig.~\ref{fig:exampletree}.

A program's fitness is determined by running it on a given set of inputs and evaluating the output.
Mutation can involve randomly substituting individual elements of a tree or replacing whole sub-trees with new randomly generated sub-trees; the latter importantly allows the tree to grow if a small sub-tree is replaced by a larger one. Crossover can involve replacing a sub-tree of one tree with a sub-tree from the other tree.

\funsearch \cite{funsearch} can be thought of as a genetic programming algorithm with a few atypical features:
\begin{itemize}
\item Instead of representing programs as trees, it represents programs as text, specifically as \Python code.
\item Instead of mutation and crossover based on elements of trees, it uses a Large Language Model (LLM). Specifically, it uses a pretrained language model, finetuned for code generation (e.g.\ Copilot \cite{copilot}, Codey \cite{codey}, CodeLlama \cite{codellama}, CodeStral \cite{codestral}). The LLM is prompted with text from two programs from the population ordered by fitness and labeled \texttt{v0} and \texttt{v1} as well as an incomplete function labeled \texttt{v2}. It then completes the function labeled \texttt{v2}, which is entered into the population subject to selection. This both introduces random variation, as the LLM will typically not reproduce the functions \texttt{v0} or \texttt{v1} verbatim, and a kind of crossover, as the function \texttt{v2} generated will be influenced by which functions \texttt{v0} and \texttt{v1} the LLM is presented with. 
\item As the code generated can contain arbitrary \Python functions, it is imperative that evaluation of programs in \funsearch take place in an appropriate sandbox environment, as these functions will not always compile or run, and may have adverse consequences for the system if they do. 
\end{itemize}
Of the features described earlier, \funsearch also uses Boltzmann selection and islands, with the number of islands being ten in the public implementation we make use of.

While we in practice want to minimize the time spent on the IBP reduction of a set of Feynman integrals to master integrals, the number of seeds and thus IBP equations used to solve the system provides an easier-to-measure proxy for this quantity. Within the context of genetic algorithms, such a minimization problem is typically formulated as a maximization problem for the fitness. 
Throughout this paper, we determine the fitness $f$ of an attempt to generate a seeding strategy as follows.
If $N_S$ is the number of seeds selected by a seeding strategy and $N_R$ is the number of seeds in the rectangular IBP system we allow the seeding strategy to select from, then the fitness is 
\begin{itemize}
\item $f=-N_S$ if the seeds solve the system,
\item $f=-N_S-N_R$ if the seeds do not solve the system for all algorithms except for our initial attempt in section~\ref{sec:ga}, where we take $f=-N_S-N_R-1$.
\end{itemize}
If the seeds successfully solve the IBP system in terms of a minimal basis of master integrals, the strategy is assigned a fitness score equal to minus the number of seeds used, thus incentivizing strategies that select a small number of seeds. However, if the seeds do not solve the system, the score receives an extra penalty of $N_R$.
This is because in principle if one has failed to solve the system one needs to try again with a list of seeds that we already know can solve the system, in this case the rectangular system.
(For one of the algorithms we try, we moreover punish failure to solve a bit more strongly than successfully solving the system with the rectangular system, hence the $-1$.)

In some of the runs of strongly typed genetic programming, we also experimented with penalizing lists of zero seeds more strongly: such systems are easy to generate 
 and they score better than all other strategies that fail to solve the system.
  Thus for some runs  
  we doubled the penalty for lists of zero size in order to avoid this local maximum.
 In practice, this did not appear to make a significant difference in performance.

\section{Initial Attempt: Genetic Algorithm}
\label{sec:ga}

We began with a genetic algorithm that used very little information about the specifics of the problem, to provide a baseline and check whether the application of more sophisticated methods is required.

Suppose that we follow rectangular seeding and end up with a list of $N_R$ seed integrals. We can specify a subset of seeds with a binary vector of length $N_R$, including only the relations generated by seeds corresponding to a $1$ in the vector. 
We performed a genetic algorithm on vectors of this form, generating an initial population where each entry has a 50\% chance to be $0$ or $1$. 

We made sure that the vector with highest fitness was always kept and otherwise used roulette selection with probability proportional to $N_R+f$, with a 5\% chance to mutate a random entry of each vector and a crossover operation that splices complementary pieces of two vectors together at a random point. Pairs  of vectors were subject to crossover 90\% of the time and remained unchanged the remaining 10\%.

For the genetic algorithm in this section, we began with rectangular systems with $s_\textrm{max}=3$ and either $r_\textrm{max}=6$ or $r_\textrm{max}=7$, which have 6,764 and 14,588 seeds, respectively. The genetic algorithm described in this section performs relatively poorly at this task, but still manages to make progress. Typically, an initial population with between 100 and 500 random vectors will find one that successfully solves the system with half the number of seeds as in the rectangular system. Subsequent generations generally lead to much less progress, though, with 100 generations only able to cut 500 additional seeds from the total and appearing to slow down with additional generations. Running 100 generations on a single CPU%
\footnote{We used either an AMD EPYC 7F72 or an Intel Xeon CPU E5-2698 v4 @ 2.20GHz, depending on the run.}
 took about 24 hours. We also tried beginning with a smaller system of seeds already included in the population, namely one that followed improved seeding with $s_\textrm{max}=3, r_\textrm{max}=6, d_\textrm{max}=1$, and $l=4$, but in this case the algorithm never found any better solutions.

Given the lack of success of this initial attempt with a classic genetic algorithm, it is warranted to employ more advanced versions of genetic algorithms, which we will do in the subsequent sections.

\section{Exploration with Funsearch}
\label{sec:funsearch}

The authors of \funsearch recommend using it in situations where it is unclear how to design a genetic algorithm to take advantage of the structure of a problem, as a tool for exploration~\cite{funsearch}. This is precisely how we will use it here.

Specifically, we built off of the fork of \funsearch in ref.~\cite{funsearchfork} which implements two features left off of the initial authors' public implementation, namely sandboxing (via containerization, which can be done via Podman~\cite{podman} or Docker~\cite{docker}) and calls to the LLM (implemented via the \texttt{llm} package \cite{llm}). We modified this code slightly, both to use an alternate containerization software available on our local cluster (Apptainer~\cite{apptainer,singularity}) and to call \Kira outside of the container to keep the container environments lightweight. We use Code Llama 7B~\cite{codellama}, a light-weight model trained for code completion, as our LLM.

To use \funsearch, one must specify an evaluation function and an initial function titled \texttt{priority} which will be included in the initial prompts to the LLM labeled \texttt{priority\_v0}. We use essentially the same evaluation function as described in the previous section, with the exception of the extra $-1$ penalty for failing to solve the system which we did not find to be necessary here. 

As above, we begin with a rectangular system, this time specifically with $r_\textrm{max}=7$. We then use the \texttt{priority} functions generated by the LLM to choose which seeds to use based on the list of $a_i$ for that seed.

We tried several initial \texttt{priority} functions. In practice, we found the code in fig.~\ref{fig:priority_prompt} to be the most successful.
\begin{figure}[tp]
\begin{verbatim}
def priority(a_list: list[int]) -> bool:
  """Decides whether to include the seed a_list in the ibp system.
     Returns True or False."""

  len_alist=len(a_list)

  #Number of propagators, which are entries in a_list greater than zero
  num_props=sum(map(lambda x: 1 if x>0 else 0,a_list))

  #Numbers of numerators, which are entries in a_list less than zero
  numerators=sum(map(lambda x: 1 if x<0 else 0,a_list))

  #Dots, the sum of all entries in a_list greater than one
  dots=sum(map(lambda x: x-1 if x>1 else 0,a_list))

  #The simplest choice: if there is more than one dot, exclude the seed
  #else include it
  if dots>1:
    return False
  else:
    return True
\end{verbatim}
\caption{Priority function used for the initial prompt to \funsearch, corresponding to a golden rule system with $d_\textrm{max}=1$.}
\label{fig:priority_prompt}
\end{figure}
Note that here comments matter: the LLM is given this function as a prompt including comments. As such, it will be biased towards generating code that would typically have comments of this sort, for example code that defines similar variables. This initial priority function corresponds to a golden rule system with $d_\textrm{max}=1$ with 2,148 seeds.

While the initial genetic algorithm found a better solution fairly quickly and then advanced steadily but slowly, \funsearch advanced in large jumps. We used \funsearch's default settings, which resulted in a much smaller population but, because this meant fewer evaluations of new solutions, much faster generations. It fairly quickly found a solution with 444 seeds, and after 1000 generations had been able to find a solution with 214 seeds, shown in fig.~\ref{fig:funsearch_moderate_result}. Using the same CPU resources of the previous section along with an Nvidia A100 or V100 GPU, this took about 16 hours. Running 1400 more generations found a solution with 92 seeds, shown in fig.~\ref{fig:funsearch_improved_seeding_result}, taking another 22 hours. This solution turned out to be equivalent to the improved seeding strategy with $d_\textrm{max}=0$ and $l=4$. Finally, running for an additional 24 hours with 1400 more generations found a solution with 88 seeds, shown in fig.~\ref{fig:funsearch_best_seeding_result}. It achieved this solution by imposing the same conditions as improved seeding, with the extra condition that the number of propagators be four or more, $t\geq4$, thus excluding seeds with three propagators which would otherwise be included in the system. While we do not know if this strategy is only valid due to a quirk of the problem we are considering, it is the case that it results in a smaller number of seeds than any seeding strategy currently on the market would provide.

\begin{figure}[tbp]
\begin{verbatim}
def priority(a_list: list[int]) -> bool:
  """Decides whether to include the seed a_list in the ibp system.
     Returns True or False."""
  #The number of negative entries in a_list
  num_negs=sum(map(lambda x: 1 if x<0 else 0,a_list))
  
  if num_negs>1:
    return False
    
  #The number of dots in a_list
  dots=sum(map(lambda x: x-1 if x>1 else 0,a_list))
  
  if dots>0:
    return False
    
  even_numbers=sum(map(lambda x: 1 if x%2==0 else 0,a_list))
  
  if even_numbers>4:
    return False
    
  #The number of positive entries in a_list
  num_props=sum(map(lambda x: 1 if x>0 else 0,a_list))
  
  if num_props+num_negs<4:
    return False
    
  return True  
\end{verbatim}
\caption{A function generated by \funsearch which gives 214 seeds for our test case.}
\label{fig:funsearch_moderate_result}
\end{figure}

\begin{figure}[tbp]
\begin{verbatim}
def priority(a_list: list[int]) -> bool:
  """Decides whether to include the seed a_list in the ibp system.
     Returns True or False."""\n  #The number of dots in a_list
  dots=sum(map(lambda x: x-1 if x>1 else 0,a_list))
  
  if dots>0:
    return False
    
  #The number of positive entries in a_list
  num_props=sum(map(lambda x: 1 if x>0 else 0,a_list))
  
  if num_props<2:
    return False
    
  #Number of nonzero elements in a_list
  nz=sum(a_list)
  
  if nz<3:
    return False
    
  if nz>8:
    return False
    
  #Number of elements less than 1 in a_list
  n1=sum(map(lambda x: 1 if x<1 else 0,a_list))
  
  if n1>4:
    return False
    
  return True
\end{verbatim}
\caption{A function generated by \funsearch which gives 92 seeds for our test case, equivalent to improved seeding with $d_\textrm{max}=0$ and $l=4$.
}
\label{fig:funsearch_improved_seeding_result}
\end{figure}

\begin{figure}[tbp]{\scriptsize
\begin{verbatim}
def priority(a_list: list[int]) -> bool:
  """Decides whether to include the seed a_list in the ibp system.
     Returns True or False."""
  if len(a_list) < 4:
    return False
    
  dots=sum(map(lambda x: x-1 if x>1 else 0,a_list))
  
  if dots>0:
    return False
    
  #The number of positive entries in a_list
  num_props=sum(map(lambda x: 1 if x>0 else 0,a_list))
  
  if num_props<2:
    return False
    
  #Number of nonzero elements in a_list
  nz=sum(a_list)
  
  if nz<3:
    return False
    
  if nz>8:
    return False
    
  #Number of elements less than 1 in a_list
  n1=sum(map(lambda x: 1 if x<1 else 0,a_list))
  
  if n1>3:
    return False
    
  #The number of entries in a_list that are less than 1/2
  n12=sum(map(lambda x: 1 if x<1/2 else 0,a_list))
  
  if n12>3:
    return False
    
  #Number of entries that are less than 1/4
  n14=sum(map(lambda x: 1 if x<1/4 else 0,a_list))
  
  if n14>3:
    return False
    
  #Number of elements less than 1/8
  n18=sum(map(lambda x: 1 if x<1/8 else 0,a_list))
  
  if n18>3:
    return False
    
  #Number of elements less than 1/16
  n116=sum(map(lambda x: 1 if x<1/16 else 0,a_list))
  
  if n116>3:
    return False
    
  return True
\end{verbatim} }
\caption{A function generated by \funsearch which gives 88 seeds, thus performing better than the improved seeding strategy for our test case.}
\label{fig:funsearch_best_seeding_result}
\end{figure}

We note here that the solutions found by \funsearch differ from what a human programmer would propose in several ways. There are several lines of code that simply have no effect in our test case: for example, the code in fig.~\ref{fig:funsearch_improved_seeding_result} demands that the number of propagators is two or greater, but this is already ensured by excluding trivial sectors, while the code in fig.~\ref{fig:funsearch_best_seeding_result} has several lines imposing conditions on $a_i$ less than a fractional number, which as all $a_i$ are integer are equivalent to a condition on $a_i$ less than one. Others impose unusual conditions that happen to restrict the list of seeds but are probably not generally useful outside of this context, in a way that would be clear to a human familiar with the problem: for example, the demand in fig.~\ref{fig:funsearch_moderate_result} that there be no more than four even numbers in the $a_i$. Others still are described misleadingly: the code in fig.~\ref{fig:funsearch_improved_seeding_result} and in its ``descendant'' in fig.~\ref{fig:funsearch_best_seeding_result} define a variable \texttt{nz} that they describe as the number of non-zero elements in \texttt{a\_list}, but actually define the number as the sum of $a_i$, which as mentioned in section~\ref{subsec:ibp} can be constrained to replicate the improved seeding strategy in cases without dots such as this one.

We were curious to what extent the results we obtained were influenced by the level of detail present in our prompt, so we investigated what happens when we use a prompt with an essentially empty priority function, shown in fig.~\ref{fig:simplest_prompt}. This initial function results in a rectangular system with 14,588 seeds. After running for a total{\parfillskip=0pt\par}
\FloatBarrier
\begin{figure}[tbp]
\begin{verbatim}
def priority(a_list: list[int]) -> bool:
  """Decides whether to include the seed a_list in the ibp system.
     Returns True or False."""

  return True
\end{verbatim}
\caption{An essentially empty priority function used to prompt \funsearch in test cases, which returns the original rectangular system.}
\label{fig:simplest_prompt}
\end{figure}
\noindent  of 34 hours \funsearch was able to find a solution with 476 seeds, shown in fig.~\ref{fig:funsearch_from_simplest_result}. Further runs from this point did not find substantial improvements, suggesting that without more extensive input \funsearch plateaus here. The resulting code is quite strange, with many redundant or irrelevant conditions on counts of ones and zeroes described in repetitive and for the most part useless ways and some odd conditions on adjacent entries.

\begin{figure}[tbp]{\footnotesize
\begin{verbatim}
def priority(a_list: list[int]) -> bool:
  """Decides whether to include the seed a_list in the ibp system.
     Returns True or False."""
  # If there are too many 1's, we can't do anything.
  if a_list.count(1) > 10:
    return False
    
  # If there are too many 0's, we can't do anything.
  if a_list.count(0) > 10:
    return False
    
  # If there are too many 1's and 0's, we can't do anything.
  if a_list.count(1) + a_list.count(0) > 12:
    return False
    
  # Check if there are enough 1's and 0's.
  if a_list.count(1) + a_list.count(0) >= 6:
    return True
    
  # If there are not enough 1's and 0's, we can't do anything.
  if a_list.count(1) + a_list.count(0) <= 4:
    return False
    
  # Check if we are in a trivial solution.
  for i in range(len(a_list) - 1):
    if a_list[i] == 0 and a_list[i + 1] == 0:
        return False
        
  # Check if we are in a trivial solution.
  for i in range(len(a_list) - 1):
    if a_list[i] == 1 and a_list[i + 1] == 1:
        return False
        
  # Find the number of 1's and 0's.
  ones = a_list.count(1)
  zeros = a_list.count(0)
  
  # If there are not enough 1's, we can't do anything.
  if ones < 3:
    return False
    
  # Find the number of 1's in groups of 3 or more.
  groups = 0
  
  # If there are too many groups, we can't do anything.
  if groups > 4:
    return False
\end{verbatim} }
\caption{The best priority function found by \funsearch from the essentially empty prompt in fig~\ref{fig:simplest_prompt}, resulting in 476 seeds.}
\label{fig:funsearch_from_simplest_result}
\end{figure}

In general, \funsearch's best solutions suggest that in addition to the constraints usually imposed during seeding that we discussed in subsection~\ref{subsec:ibp} (such as restrictions on $t,r,d$, and $s$) we should also consider restrictions on the full sum of $a_i$ and on the count of entries equal to 1 or 0. We will use these insights in the next section to more efficiently find our best seeding strategy using a more traditional approach to genetic programming.

\section{Improved Heuristics via Strongly Typed Genetic Programming}
\label{sec:gp}

Based on the exploration in the previous section, we now use genetic programming to evolve seeding strategies, which places more restrictions on what kind of conditions can arise compared to \funsearch.
In genetic programming, one must specify
\begin{itemize}
\item the arguments of the program to evolve,
\item a list of primitive elements, which are functions that may be included in the program,
\item optionally, a list of terminal elements, elements which are not functions or arguments, and
\item as in genetic algorithms in general, one also needs an evaluation function.
\end{itemize}
We will specifically use strongly typed genetic programming, in which our primitive elements have specified data types for input and output which constrain which elements can follow each other.
We employ the implementation of strongly typed genetic programming in the \DEAP package~\cite{deap}. Specifically, we use that package's \texttt{eaSimple} algorithm, with tournament selection using three-member tournaments.

As above, the program we want to evolve will be a function to decide whether or not to include a seed with a given list of $a_i$ in the system, returning \texttt{True} if the seed is to be included and \texttt{False} if not. In principle, one has a large number of choices for how to specify the arguments of the program and the list of primitive elements consistent with this goal. One intuitive choice would be to let the arguments be the individual $a_i$, and then have primitive elements that include simple operations on integers (greater than $>$, less than $<$, equal to $=$, sum $+$ and difference $-$), boolean operations (and, or, not), and terminal elements including simple integers (say, between $-10$ and $+10$). However, we find that this choice performs poorly, often not even finding a valid solution.

We can gain more insight by looking at the solutions which were successful in \funsearch. Typically, these solutions did not involve imposing conditions on the individual $a_i$. Instead, they used a relatively small list of variables constructed out of the $a_i$, including sums of the full list, total propagator and numerator powers, number of dots, number of propagators, and number of zeros. Inspired by this, we chose the following lists of arguments, primitives, and terminal elements for our genetic programming.

As arguments, we chose
\begin{itemize}
\item \texttt{sum\_gt\_0}: the sum of all $a_i$ greater than zero,
\item \texttt{sum\_gt\_1}: the sum of all $a_i$ greater than one
\item \texttt{minus\_sum\_lt\_0}: minus the sum of all $a_i$ less than zero,
\item \texttt{sum\_all}: the sum of all $a_i$,
\item \texttt{count\_gt\_0}: the number of $a_i$ greater than zero,
\item \texttt{count\_gt\_1}: the number of $a_i$ greater than one,
\item \texttt{count\_lt\_0}: the number of $a_i$ less than zero,
\item \texttt{count\_eq\_0}: the number of $a_i$ equal to zero,
\item \texttt{count\_eq\_1}: the number of $a_i$ equal to one, and 
\item \texttt{count\_all}: the length of the list of $a_i$.
\end{itemize}
As primitives, we chose 
\begin{itemize}
\item \texttt{and\_}: and, which takes two booleans returning a boolean,
\item \texttt{gt}: greater than, which takes two numbers returning a boolean,
\item \texttt{lt}: less than, which takes two numbers and returns a boolean,
\item \texttt{eq}: equal to, which takes two numbers and returns a boolean,
\item \texttt{add}: addition, which takes two numbers returning a number, and 
\item \texttt{sub}: subtraction, which takes two numbers returning a number.
\end{itemize}
Finally, as the terminal elements we chose
\begin{itemize}
\item \texttt{True},
\item \texttt{0},
\item \texttt{r\_max},
\item \texttt{s\_max},
\item and the integers between $-10$ and $+10$.
\end{itemize}

 When we construct the initial population, we use \DEAP's function \texttt{genHalfAndHalf} to generate a random valid tree with depth between 3 and 5.  \texttt{genHalfAndHalf} has a 50\% chance of generating a tree where each leaf has the same depth, and a 50\% chance of generating a tree where each leaf can have different depth. When building trees, \DEAP first determines whether a node will be terminal, then chooses uniformly between the appropriate bullet points above, choosing an argument or terminal element for terminal nodes and one of the other primitives for non-terminal nodes. 
 We thus include \texttt{True} as a primitive element because \DEAP requires there to be a terminal element of each type, and we list \texttt{0} separately in addition to being included in the integers as constraining something (for example dots) to zero should be a frequent move, so \DEAP would have an equal chance of completing a branch of a tree with \texttt{0} specifically and completing it with a random uniformly chosen integer between $-10$ and $+10$ inclusive.
 
 For each generation, two individuals have a 50\% chance to crossover, replacing a randomly chosen sub-tree of one with a randomly chosen sub-tree of another, and a 10\% chance to mutate, replacing a randomly chosen sub-tree with a freshly generated sub-tree of depth 0 to 2. Both operations are restricted to never generate individuals with depth greater than 17.

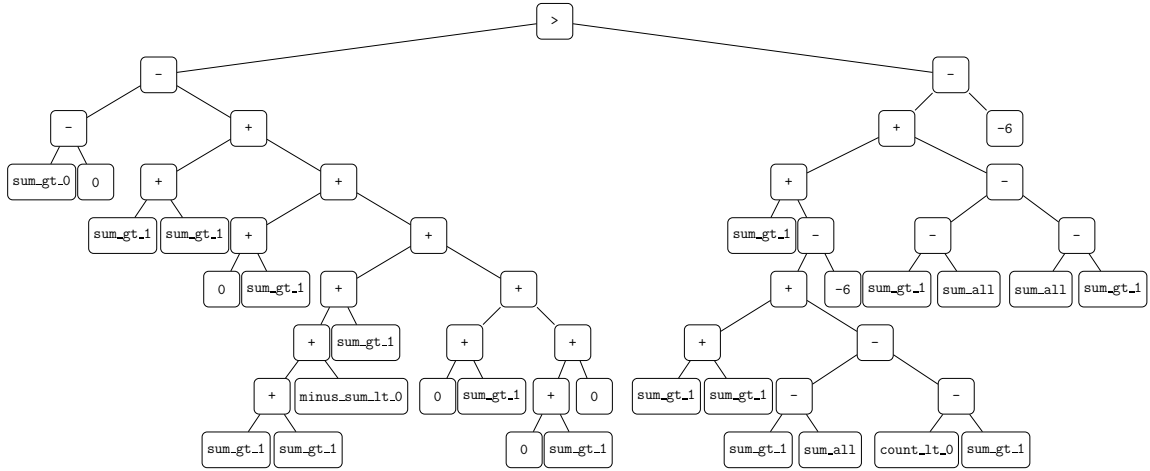
\begin{figure}[tp]
\resizebox{\textwidth}{!}{
\begin{tikzpicture}[node distance=1.5cm,
    every node/.style={fill=white, font=\sffamily}, align=center]

\node(greater1) [base] {>};

\node(sub1) [base,below of=greater1, xshift=-11cm] {-};

\draw(greater1) -- (sub1);

\node(sub2) [base,below of=sub1, xshift=-2.5cm] {-};

\draw(sub1) -- (sub2);

\node(sumgt01) [base,below of=sub2, xshift=-0.75cm] {sum\_gt\_0};

\draw(sub2) -- (sumgt01);

\node(zero1) [base,below of=sub2, xshift=0.75cm] {0};

\draw(sub2) -- (zero1);

\node(add1) [base,below of=sub1, xshift=2.5cm] {+};

\draw(sub1) -- (add1);

\node(add2) [base,below of=add1, xshift=-2.5cm] {+};

\draw(add1) -- (add2);

\node(sumgt11) [base,below of=add2, xshift=-1cm] {sum\_gt\_1};

\draw(add2) -- (sumgt11);

\node(sumgt12) [base,below of=add2, xshift=1cm] {sum\_gt\_1};

\draw(add2) -- (sumgt12);

\node(add3) [base,below of=add1, xshift=2.5cm] {+};

\draw(add1) -- (add3);

\node(add4) [base,below of=add3, xshift=-2.5cm] {+};

\draw(add3) -- (add4);

\node(zero2) [base,below of=add4, xshift=-0.75cm] {0};

\draw(add4) -- (zero2);

\node(sumgt13) [base,below of=add4, xshift=0.75cm] {sum\_gt\_1};

\draw(add4) -- (sumgt13);

\node(add5) [base,below of=add3, xshift=2.5cm] {+};

\draw(add3) -- (add5);

\node(add6) [base,below of=add5, xshift=-2.5cm] {+};

\draw(add5) -- (add6);

\node(add7) [base,below of=add6, xshift=-0.75cm] {+};

\draw(add6) -- (add7);

\node(add8) [base,below of=add7, xshift=-1.1cm] {+};

\draw(add7) -- (add8);

\node(sumgt14) [base,below of=add8, xshift=-1cm] {sum\_gt\_1};

\draw(add8) -- (sumgt14);

\node(sumgt15) [base,below of=add8, xshift=1cm] {sum\_gt\_1};

\draw(add8) -- (sumgt15);

\node(minussumlt01) [base,below of=add7, xshift=1.1cm] {minus\_sum\_lt\_0};

\draw(add7) -- (minussumlt01);

\node(sumgt16) [base,below of=add6, xshift=0.75cm] {sum\_gt\_1};

\draw(add6) -- (sumgt16);

\node(add9) [base,below of=add5, xshift=2.5cm] {+};

\draw(add5) -- (add9);

\node(add10) [base,below of=add9, xshift=-1.5cm] {+};

\draw(add9) -- (add10);

\node(zero3) [base,below of=add10, xshift=-0.75cm] {0};

\draw(add10) -- (zero3);

\node(sumgt17) [base,below of=add10, xshift=0.75cm] {sum\_gt\_1};

\draw(add10) -- (sumgt17);

\node(add11) [base,below of=add9, xshift=1.5cm] {+};

\draw(add9) -- (add11);

\node(add12) [base,below of=add11, xshift=-0.6cm] {+};

\draw(add11) -- (add12);

\node(zero4) [base,below of=add12, xshift=-0.75cm] {0};

\draw(add12) -- (zero4);

\node(sumgt18) [base,below of=add12, xshift=0.75cm] {sum\_gt\_1};

\draw(add12) -- (sumgt18);

\node(zero5) [base,below of=add11, xshift=0.6cm] {0};

\draw(add11) -- (zero5);

\node(sub3) [base,below of=greater1, xshift=11cm] {-};

\draw(greater1) -- (sub3);

\node(add13) [base,below of=sub3, xshift=-1.5cm] {+};

\draw(sub3) -- (add13);

\node(add14) [base,below of=add13, xshift=-3cm] {+};

\draw(add13) -- (add14);

\node(sumgt19) [base,below of=add14, xshift=-0.75cm] {sum\_gt\_1};

\draw(add14) -- (sumgt19);

\node(sub4) [base,below of=add14, xshift=0.75cm] {-};

\draw(add14) -- (sub4);

\node(add15) [base,below of=sub4, xshift=-0.75cm] {+};

\draw(sub4) -- (add15);

\node(add16) [base,below of=add15, xshift=-2.4cm] {+};

\draw(add15) -- (add16);

\node(sumgt110) [base,below of=add16, xshift=-1cm] {sum\_gt\_1};

\draw(add16) -- (sumgt110);

\node(sumgt111) [base,below of=add16, xshift=1cm] {sum\_gt\_1};

\draw(add16) -- (sumgt111);

\node(sub5) [base,below of=add15, xshift=2.4cm] {-};

\draw(add15) -- (sub5);

\node(sub6) [base,below of=sub5, xshift=-2.25cm] {-};

\draw(sub5) -- (sub6);

\node(sumgt112) [base,below of=sub6, xshift=-1cm] {sum\_gt\_1};

\draw(sub6) -- (sumgt112);

\node(sumall1) [base,below of=sub6, xshift=1cm] {sum\_all};

\draw(sub6) -- (sumall1);

\node(sub7) [base,below of=sub5, xshift=2.25cm] {-};

\draw(sub5) -- (sub7);

\node(countlt0) [base,below of=sub7, xshift=-1.1cm] {count\_lt\_0};

\draw(sub7) -- (countlt0);

\node(sumgt113) [base,below of=sub7, xshift=1.1cm] {sum\_gt\_1};

\draw(sub7) -- (sumgt113);

\node(const1) [base,below of=sub4, xshift=0.75cm] {-6};

\draw(sub4) -- (const1);

\node(sub8) [base,below of=add13, xshift=3cm] {-};

\draw(add13) -- (sub8);

\node(sub9) [base,below of=sub8, xshift=-2cm] {-};

\draw(sub8) -- (sub9);

\node(sumgt114) [base,below of=sub9, xshift=-1cm] {sum\_gt\_1};

\draw(sub9) -- (sumgt114);

\node(sumall2) [base,below of=sub9, xshift=1cm] {sum\_all};

\draw(sub9) -- (sumall2);

\node(sub10) [base,below of=sub8, xshift=2cm] {-};

\draw(sub8) -- (sub10);

\node(sumall3) [base,below of=sub10, xshift=-1cm] {sum\_all};

\draw(sub10) -- (sumall3);

\node(sumgt115) [base,below of=sub10, xshift=1cm] {sum\_gt\_1};

\draw(sub10) -- (sumgt115);

\node(const2) [base,below of=sub3, xshift=1.5cm] {-6};

\draw(sub3) -- (const2);

\end{tikzpicture}
}
\caption{A tree diagram for a program generated via strongly typed genetic programming in \DEAP which achieves our best-case seeding of 88 seeds.}
\label{fig:gptree}
\end{figure}

Running with a population of 300, we found that we could find the best solution found by \funsearch, with 88 seeds, in fairly few generations, with one run finding this seeding strategy in only 18 generations while others found it in 30--40.\footnote{The longest of these runs ran for eight hours on a single CPU, with some running for less than three.} A particularly interesting function that achieved this best-case seeding is depicted in fig.~\ref{fig:gptree}.

The function in fig.~\ref{fig:gptree} may appear quite complicated, however, it can be readily seen that many of the conditions it imposes are redundant. Cleaning redundant conditions and simplifying by removing statements which are always true in the context of our example, we are left with
\begin{equation}
4\texttt{sum\_all}+\texttt{count\_lt\_0} > 15\texttt{sum\_gt\_1}+12\,.
\end{equation}
This result achieves our best-case seeding in a fairly interesting way. As the multiplier on \texttt{sum\_gt\_1} is quite high, it appearing on the right-hand size of the inequality forces this term to vanish, as it is impossible for the left-hand size to be high enough to be greater than it. This enforces $d=0$. If there is at least one numerator then we have $\sum a_i \geq 3$, which is equivalent to the condition imposed by improved seeding. If we have no numerators then we instead have $\sum a_i > 3$, which in this situation demands that there be at least four propagators, the additional condition that takes the number of seeds down to 88.

The results above were obtained with minimal tuning of the hyperparameters, such as the probabilities for different actions during the generation of the trees, the probability for crossover and mutation, and the maximal depth.
It is likely that a scan over hyperparameters or the re-inclusion of individual $a_i$s with low probability leads to even smaller systems of seeds. 
We leave corresponding investigations for future work.

\section{Conclusions and Discussion}
\label{sec:conclusion}

Integration-by-parts reduction is a frequent bottle neck in state-of-the-art calculations in perturbative Quantum Field Theory, making it a crucial target for improvements. 
In this paper, we have applied machine-learning techniques to this problem, adding to a short but growing list of applications of machine learning to analytic calculations in theoretical high-energy physics \cite{Dersy:2022bym,Cai:2024znx,Cheung:2024svk,Koay:2025bmu}. 

Surprisingly simple changes to the heuristics used for seeding integration-by-parts systems can have a dramatic effect. In such an environment, the ability to try a large number of heuristics, recombining the best parts to form new ones, has great potential, reproducing in a way the experimentation that can occur within a scientific community.
We have found that, using methods from genetic algorithms, we can rediscover the latest strategy from known seeding algorithms~\cite{JohannQCDmeetsGravity,Driesse:2024xad,Guan:2024byi,Bern:2024adl} and even modestly improve on it. Knowing very little about the kinds of methods we needed, we could find these improvements via the methodology of \funsearch~\cite{funsearch}, generating code with a Large Language Model. With a bit of inspiration from these results, we could use a more classic type of strongly typed genetic programming instead, leading to much faster convergence.

In this paper, we have provided a proof of principle that genetic programming can be used to improve seeding strategies, using a simple two-loop Feynman integral as a benchmark.
In the future, it would be interesting to consider also larger sets of more complicated Feynman integrals. In contrast to other machine-learning methods, the strategies produced by \funsearch and strongly typed genetic programming are fully interpretable. Considering a range of different integrals thus promises to reveal fully general strategies that can be included in future IBP software.
Complementarily, we could also imagine the machine-learning techniques being incorporated into IBP software to find optimal seeding strategies tailored to a given problem and corresponding set of integrals. As current methods rely on solving the same IBP system many times on different kinematic points for rational reconstruction, it should be possible to use genetic programming over the course of a reconstruction, optimizing the seed list while evaluating at different kinematic points 
 so as to make the subsequent evaluations faster.

In the genetic algorithms used in this work, we began with a large system of seeds and applied a filter. One could imagine progressing in the opposite way, beginning with a small list of seeds containing the integrals of interest and learning how to expand efficiently to solve the full system. We will explore this idea in future work~\cite{in_progress}.

\section*{Acknowledgements}

We thank 
Justin Berman,
Fran\c{c}ois Charton,
Jordan Ellenberg,
Garrett Merz, 
Maja Rudolph
 and 
Johann Usovitsch
for fruitful discussions,
Baptiste Rozière and Alexander Smirnov for communication,
Fran\c{c}ois Charton and Johann Usovitsch for comments on the manuscript
 as well as 
Cynthia Rodr{\'i}guez for initial collaboration.
Parts of the computations done for this project were performed on the UCloud interactive HPC system, which is managed by the eScience Center at the University of Southern Denmark. Other parts were performed on SCIENCE AI Centre's GPU cluster at the University of Copenhagen.
The work of MvH and MW was supported by the research grant 00025445 from Villum Fonden.  MW was further supported by the Sapere Aude: DFF-Starting Grant 4251-00029B. 
MW moreover acknowledges the warm hospitality of the Data Science Institute, University of Wisconsin.

\appendix

\section{Integration-by-Parts identities for the Benchmark Integral}
\label{sec:fullrelations}

In this appendix, we provide the explicit form of the IBP identities for the integral depicted in fig~\ref{fig:triangle}, which is the example we use as a benchmark for all machine-learning approaches in this paper.

In total, there are 8 IBP identities for general indices $a_i$:
\begin{align}
&(D-a_3-a_7-a_4-2a_1)I(a_1,a_2,a_3,a_4,a_5,a_6,a_7)\nonumber\\*&-a_3I(a_1-1,a_2,a_3+1,a_4,a_5,a_6,a_7)+a_3I(a_1,a_2-1,a_3+1,a_4,a_5,a_6,a_7)\nonumber\\*&+a_7m_4I(a_1,a_2,a_3,a_4,a_5,a_6,a_7+1)-a_4I(a_1-1,a_2,a_3,a_4+1,a_5,a_6,a_7)\nonumber\\*&-a_7I(a_1-1,a_2,a_3,a_4,a_5,a_6,a_7+1)+a_4I(a_1,a_2,a_3,a_4+1,a_5,a_6,a_7)=0\,,\\
&(D-2a_2-a_6-a_3-a_5)I(a_1,a_2,a_3,a_4,a_5,a_6,a_7)\nonumber\\*&+a_3I(a_1-1,a_2,a_3+1,a_4,a_5,a_6,a_7)-a_3I(a_1,a_2-1,a_3+1,a_4,a_5,a_6,a_7)\nonumber\\*&+a_6I(a_1,a_2,a_3,a_4,a_5,a_6+1,a_7)-a_5I(a_1,a_2-1,a_3,a_4,a_5+1,a_6,a_7)\nonumber\\*&-a_6I(a_1,a_2-1,a_3,a_4,a_5,a_6+1,a_7)+m_4a_5I(a_1,a_2,a_3,a_4,a_5+1,a_6,a_7)=0\,,\\
&(-a_4+a_1)I(a_1,a_2,a_3,a_4,a_5,a_6,a_7)-a_1I(a_1+1,a_2,a_3,a_4-1,a_5,a_6,a_7)\nonumber\\*&+a_1I(a_1+1,a_2,a_3,a_4,a_5,a_6,a_7)+a_3I(a_1-1,a_2,a_3+1,a_4,a_5,a_6,a_7)\nonumber\\*&-a_3I(a_1,a_2-1,a_3+1,a_4,a_5,a_6,a_7)-a_3I(a_1,a_2,a_3+1,a_4-1,a_5,a_6,a_7)\nonumber\\*&+a_3I(a_1,a_2,a_3+1,a_4,a_5,a_6-1,a_7)+a_4I(a_1-1,a_2,a_3,a_4+1,a_5,a_6,a_7)\nonumber\\*&+(-a_7m_3^2+2a_7+a_7m_4)I(a_1,a_2,a_3,a_4,a_5,a_6,a_7+1)\nonumber\\*&-a_4I(a_1,a_2,a_3,a_4+1,a_5,a_6,a_7)+a_7I(a_1-1,a_2,a_3,a_4,a_5,a_6,a_7+1)\nonumber\\*&-a_7I(a_1,a_2,a_3,a_4-1,a_5,a_6,a_7+1)=0\,,\\
&(-a_2+a_6)I(a_1,a_2,a_3,a_4,a_5,a_6,a_7)+a_2I(a_1,a_2+1,a_3,a_4,a_5,a_6-1,a_7)\nonumber\\*&-a_2I(a_1,a_2+1,a_3,a_4,a_5,a_6,a_7)+a_3I(a_1-1,a_2,a_3+1,a_4,a_5,a_6,a_7)\nonumber\\*&-a_3I(a_1,a_2-1,a_3+1,a_4,a_5,a_6,a_7)-a_3I(a_1,a_2,a_3+1,a_4-1,a_5,a_6,a_7)\nonumber\\*&+a_3I(a_1,a_2,a_3+1,a_4,a_5,a_6-1,a_7)-a_5I(a_1,a_2-1,a_3,a_4,a_5+1,a_6,a_7)\nonumber\\*&+a_5I(a_1,a_2,a_3,a_4,a_5+1,a_6-1,a_7)+(m_4a_5-a_5m_3^2)I(a_1,a_2,a_3,a_4,a_5+1,a_6,a_7)\nonumber\\*&-a_6I(a_1,a_2-1,a_3,a_4,a_5,a_6+1,a_7)+a_6I(a_1,a_2,a_3,a_4,a_5,a_6+1,a_7)=0\,,\\
&a_2I(a_1-1,a_2+1,a_3,a_4,a_5,a_6,a_7)+(a_2-a_3)I(a_1,a_2,a_3,a_4,a_5,a_6,a_7)\nonumber\\*&-a_2I(a_1,a_2+1,a_3-1,a_4,a_5,a_6,a_7)-a_3I(a_1-1,a_2,a_3+1,a_4,a_5,a_6,a_7)\nonumber\\*&+a_3I(a_1,a_2-1,a_3+1,a_4,a_5,a_6,a_7)-a_6I(a_1,a_2,a_3,a_4,a_5,a_6+1,a_7)\nonumber\\*&+2a_5I(a_1-1,a_2,a_3,a_4,a_5+1,a_6,a_7)+a_5I(a_1,a_2-1,a_3,a_4,a_5+1,a_6,a_7)\nonumber\\*&-a_5I(a_1,a_2,a_3-1,a_4,a_5+1,a_6,a_7)-a_5I(a_1,a_2,a_3,a_4,a_5+1,a_6,a_7-1)\nonumber\\*&+(m_4a_5)I(a_1,a_2,a_3,a_4,a_5+1,a_6,a_7)+a_6I(a_1,a_2-1,a_3,a_4,a_5,a_6+1,a_7)\nonumber\\*&-a_6I(a_1,a_2,a_3-1,a_4,a_5,a_6+1,a_7)+a_6I(a_1,a_2,a_3,a_4-1,a_5,a_6+1,a_7)=0\,,\\
&(-a_3+a_1)I(a_1,a_2,a_3,a_4,a_5,a_6,a_7)+a_1I(a_1+1,a_2-1,a_3,a_4,a_5,a_6,a_7)\nonumber\\*&-a_1I(a_1+1,a_2,a_3-1,a_4,a_5,a_6,a_7)+a_3I(a_1-1,a_2,a_3+1,a_4,a_5,a_6,a_7)\nonumber\\*&-a_3I(a_1,a_2-1,a_3+1,a_4,a_5,a_6,a_7)+(a_7m_4)I(a_1,a_2,a_3,a_4,a_5,a_6,a_7+1)\nonumber\\*&+a_4I(a_1-1,a_2,a_3,a_4+1,a_5,a_6,a_7)-a_4I(a_1,a_2,a_3-1,a_4+1,a_5,a_6,a_7)\nonumber\\*&+a_4I(a_1,a_2,a_3,a_4+1,a_5,a_6-1,a_7)-a_4I(a_1,a_2,a_3,a_4+1,a_5,a_6,a_7)\nonumber\\*&+a_7I(a_1-1,a_2,a_3,a_4,a_5,a_6,a_7+1)+2a_7I(a_1,a_2-1,a_3,a_4,a_5,a_6,a_7+1)\nonumber\\*&-a_7I(a_1,a_2,a_3-1,a_4,a_5,a_6,a_7+1)-a_7I(a_1,a_2,a_3,a_4,a_5-1,a_6,a_7+1)=0\,,\\
&a_2I(a_1,a_2+1,a_3,a_4,a_5-1,a_6,a_7)-a_2I(a_1,a_2+1,a_3,a_4,a_5,a_6-1,a_7)\nonumber\\*&+(a_2-a_2m_4)I(a_1,a_2+1,a_3,a_4,a_5,a_6,a_7)-2a_3I(a_1-1,a_2,a_3+1,a_4,a_5,a_6,a_7)\nonumber\\*&+a_3I(a_1,a_2,a_3+1,a_4-1,a_5,a_6,a_7)+a_3I(a_1,a_2,a_3+1,a_4,a_5-1,a_6,a_7)\nonumber\\*&-a_3I(a_1,a_2,a_3+1,a_4,a_5,a_6-1,a_7)+a_3I(a_1,a_2,a_3+1,a_4,a_5,a_6,a_7-1)\nonumber\\*&-2a_3m_4I(a_1,a_2,a_3+1,a_4,a_5,a_6,a_7)+(a_5-a_6)I(a_1,a_2,a_3,a_4,a_5,a_6,a_7)\nonumber\\*&-a_5I(a_1,a_2,a_3,a_4,a_5+1,a_6-1,a_7)+a_5m_3^2I(a_1,a_2,a_3,a_4,a_5+1,a_6,a_7)\nonumber\\*&+a_6I(a_1,a_2,a_3,a_4,a_5-1,a_6+1,a_7)-a_6m_3^2I(a_1,a_2,a_3,a_4,a_5,a_6+1,a_7)=0\,,\\&(a_7+a_4-2a_1)I(a_1,a_2,a_3,a_4,a_5,a_6,a_7)+a_1I(a_1+1,a_2,a_3,a_4-1,a_5,a_6,a_7)\nonumber\\*&+a_1I(a_1+1,a_2,a_3,a_4,a_5,a_6,a_7-1)-(a_1+m_4a_1)I(a_1+1,a_2,a_3,a_4,a_5,a_6,a_7)\nonumber\\*&-2a_3I(a_1-1,a_2,a_3+1,a_4,a_5,a_6,a_7)+a_3I(a_1,a_2,a_3+1,a_4-1,a_5,a_6,a_7)\nonumber\\*&+a_3I(a_1,a_2,a_3+1,a_4,a_5-1,a_6,a_7)-a_3I(a_1,a_2,a_3+1,a_4,a_5,a_6-1,a_7)\nonumber\\*&+a_3I(a_1,a_2,a_3+1,a_4,a_5,a_6,a_7-1)-2a_3m_4I(a_1,a_2,a_3+1,a_4,a_5,a_6,a_7)\nonumber\\*&-2a_4I(a_1-1,a_2,a_3,a_4+1,a_5,a_6,a_7)+(a_7m_3^2-2a_7)I(a_1,a_2,a_3,a_4,a_5,a_6,a_7+1)\nonumber\\*&+a_4I(a_1,a_2,a_3,a_4+1,a_5,a_6,a_7-1)+(a_4m_3^2-2m_4a_4)I(a_1,a_2,a_3,a_4+1,a_5,a_6,a_7)\nonumber\\*&-2a_7I(a_1-1,a_2,a_3,a_4,a_5,a_6,a_7+1)+a_7I(a_1,a_2,a_3,a_4-1,a_5,a_6,a_7+1)=0\,.
\end{align}

Moreover, there is a single LI relation, namely
\begin{align}
&a_3(3-m_2^2+m_3^2)I(a_1-1,a_2,a_3,a_4+1,a_5,a_6,a_7)\nonumber\\*&+(a_3(m_2^2-m_3^2-1)+a_4(1+m_3^2-m_2^2)+a_5(m_2^2-m_3^2-1)-a_6(1+m_2^2+m_3^2))\times\nonumber\\*&\quad\quad I(a_1,a_2,a_3,a_4,a_5,a_6,a_7)\nonumber\\*&+a_3(1-m_2^2+3m_3^2)I(a_1,a_2,a_3,a_4+1,a_5,a_6,a_7)\nonumber\\*&+a_6(m_3^2m_2^2-3m_3^2-m_3^4)I(a_1,a_2,a_3,a_4,a_5,a_6,a_7+1)\nonumber\\*&+2a_6m_3^2I(a_1,a_2,a_3,a_4-1,a_5,a_6,a_7+1)-2a_3I(a_1,a_2,a_3,a_4+1,a_5,a_6,a_7-1)\nonumber\\*&-2m_3^2a_4I(a_1,a_2,a_3,a_4,a_5+1,a_6-1,a_7)\nonumber\\*&+a_4(m_3^2+m_2^2-1)I(a_1,a_2-1,a_3,a_4,a_5+1,a_6,a_7)\nonumber\\*&+a_6(m_2^2-3m_3^2-1)I(a_1-1,a_2,a_3,a_4,a_5,a_6,a_7+1)\nonumber\\*&+a_4 m_3^2(1+m_2^2-m_3^2)I(a_1,a_2,a_3,a_4,a_5+1,a_6,a_7)\nonumber\\*&+a_5(1-m_2^2-m_3^2)I(a_1,a_2,a_3,a_4,a_5,a_6+1,a_7)\nonumber\\*&+2a_5I(a_1,a_2,a_3,a_4,a_5-1,a_6+1,a_7)\nonumber\\*&+a_5(m_3^2-m_2^2-1)I(a_1,a_2-1,a_3,a_4,a_5,a_6+1,a_7)=0\,.
\end{align}

\bibliography{bibliography}

\providecommand{\href}[2]{#2}\begingroup\raggedright\begin{thebibliography}{10}

\bibitem{Tkachov:1981wb}
F.V.~Tkachov, \emph{{A theorem on analytical calculability of 4-loop
  renormalization group functions}},
  \href{https://doi.org/10.1016/0370-2693(81)90288-4}{\emph{Phys. Lett. B}
  {\bfseries 100} (1981) 65}.

\bibitem{Chetyrkin:1981qh}
K.G.~Chetyrkin and F.V.~Tkachov, \emph{{Integration by Parts: The Algorithm to
  Calculate beta Functions in 4 Loops}},
  \href{https://doi.org/10.1016/0550-3213(81)90199-1}{\emph{Nucl. Phys. B}
  {\bfseries 192} (1981) 159}.

\bibitem{Laporta:2000dsw}
S.~Laporta, \emph{{High precision calculation of multiloop Feynman integrals by
  difference equations}},
  \href{https://doi.org/10.1142/S0217751X00002159}{\emph{Int. J. Mod. Phys. A}
  {\bfseries 15} (2000) 5087}
  [\href{https://arxiv.org/abs/hep-ph/0102033}{{\ttfamily hep-ph/0102033}}].

\bibitem{Driesse:2024xad}
M.~Driesse, G.U.~Jakobsen, G.~Mogull, J.~Plefka, B.~Sauer and J.~Usovitsch,
  \emph{{Conservative Black Hole Scattering at Fifth Post-Minkowskian and First
  Self-Force Order}},
  \href{https://doi.org/10.1103/PhysRevLett.132.241402}{\emph{Phys. Rev. Lett.}
  {\bfseries 132} (2024) 241402}
  [\href{https://arxiv.org/abs/2403.07781}{{\ttfamily 2403.07781}}].

\bibitem{Weinzierl:2022eaz}
S.~Weinzierl, \emph{{Feynman Integrals. A Comprehensive Treatment for Students
  and Researchers}}, UNITEXT for Physics, Springer (2022),
  \href{https://doi.org/10.1007/978-3-030-99558-4}{10.1007/978-3-030-99558-4},
  [\href{https://arxiv.org/abs/2201.03593}{{\ttfamily 2201.03593}}].

\bibitem{Anastasiou:2004vj}
C.~Anastasiou and A.~Lazopoulos, \emph{{Automatic integral reduction for higher
  order perturbative calculations}},
  \href{https://doi.org/10.1088/1126-6708/2004/07/046}{\emph{JHEP} {\bfseries
  07} (2004) 046} [\href{https://arxiv.org/abs/hep-ph/0404258}{{\ttfamily
  hep-ph/0404258}}].

\bibitem{Smirnov:2008iw}
A.V.~Smirnov, \emph{{Algorithm FIRE -- Feynman Integral REduction}},
  \href{https://doi.org/10.1088/1126-6708/2008/10/107}{\emph{JHEP} {\bfseries
  10} (2008) 107} [\href{https://arxiv.org/abs/0807.3243}{{\ttfamily
  0807.3243}}].

\bibitem{Smirnov:2023yhb}
A.V.~Smirnov and M.~Zeng, \emph{{FIRE 6.5: Feynman integral reduction with new
  simplification library}},
  \href{https://doi.org/10.1016/j.cpc.2024.109261}{\emph{Comput. Phys. Commun.}
  {\bfseries 302} (2024) 109261}
  [\href{https://arxiv.org/abs/2311.02370}{{\ttfamily 2311.02370}}].

\bibitem{vonManteuffel:2012np}
A.~von Manteuffel and C.~Studerus, \emph{{Reduze 2 - Distributed Feynman
  Integral Reduction}},  \href{https://arxiv.org/abs/1201.4330}{{\ttfamily
  1201.4330}}.

\bibitem{Lee:2012cn}
R.N.~Lee, \emph{{Presenting LiteRed: a tool for the Loop InTEgrals REDuction}},
   \href{https://arxiv.org/abs/1212.2685}{{\ttfamily 1212.2685}}.

\bibitem{Maierhofer:2017gsa}
P.~Maierh\"ofer, J.~Usovitsch and P.~Uwer, \emph{{Kira\textemdash{}A Feynman
  integral reduction program}},
  \href{https://doi.org/10.1016/j.cpc.2018.04.012}{\emph{Comput. Phys. Commun.}
  {\bfseries 230} (2018) 99}
  [\href{https://arxiv.org/abs/1705.05610}{{\ttfamily 1705.05610}}].

\bibitem{Klappert:2020nbg}
J.~Klappert, F.~Lange, P.~Maierh\"ofer and J.~Usovitsch, \emph{{Integral
  reduction with Kira 2.0 and finite field methods}},
  \href{https://doi.org/10.1016/j.cpc.2021.108024}{\emph{Comput. Phys. Commun.}
  {\bfseries 266} (2021) 108024}
  [\href{https://arxiv.org/abs/2008.06494}{{\ttfamily 2008.06494}}].

\bibitem{Peraro:2019svx}
T.~Peraro, \emph{{$\text{FiniteFlow}$: multivariate functional reconstruction
  using finite fields and dataflow graphs}},
  \href{https://doi.org/10.1007/JHEP07(2019)031}{\emph{JHEP} {\bfseries 07}
  (2019) 031} [\href{https://arxiv.org/abs/1905.08019}{{\ttfamily
  1905.08019}}].

\bibitem{Guan:2024byi}
X.~Guan, X.~Liu, Y.-Q.~Ma and W.-H.~Wu, \emph{{Blade: A package for
  block-triangular form improved Feynman integrals decomposition}},
  \href{https://arxiv.org/abs/2405.14621}{{\ttfamily 2405.14621}}.

\bibitem{JohannQCDmeetsGravity}
J.~Usovitsch, ``Improved integral reduction with kira.'' Talk at QCD meets
  Gravity at CERN, December 13th 2023. Slides available at
  \url{https://indico.cern.ch/event/1317494/contributions/5697745/attachments/2770593/4827307/Kira_QCD_meets_Gravity.pdf}.

\bibitem{Bern:2024adl}
Z.~Bern, E.~Herrmann, R.~Roiban, M.S.~Ruf, A.V.~Smirnov, V.A.~Smirnov et~al.,
  \emph{{Amplitudes, supersymmetric black hole scattering at $
  \mathcal{O}\left({G}^5\right) $, and loop integration}},
  \href{https://doi.org/10.1007/JHEP10(2024)023}{\emph{JHEP} {\bfseries 10}
  (2024) 023} [\href{https://arxiv.org/abs/2406.01554}{{\ttfamily
  2406.01554}}].

\bibitem{Gluza:2010ws}
J.~Gluza, K.~Kajda and D.A.~Kosower, \emph{{Towards a Basis for Planar Two-Loop
  Integrals}}, \href{https://doi.org/10.1103/PhysRevD.83.045012}{\emph{Phys.
  Rev. D} {\bfseries 83} (2011) 045012}
  [\href{https://arxiv.org/abs/1009.0472}{{\ttfamily 1009.0472}}].

\bibitem{Wu:2023upw}
Z.~Wu, J.~Boehm, R.~Ma, H.~Xu and Y.~Zhang, \emph{{NeatIBP 1.0, a package
  generating small-size integration-by-parts relations for Feynman integrals}},
  \href{https://doi.org/10.1016/j.cpc.2023.108999}{\emph{Comput. Phys. Commun.}
  {\bfseries 295} (2024) 108999}
  [\href{https://arxiv.org/abs/2305.08783}{{\ttfamily 2305.08783}}].

\bibitem{Mastrolia:2018uzb}
P.~Mastrolia and S.~Mizera, \emph{{Feynman Integrals and Intersection Theory}},
  \href{https://doi.org/10.1007/JHEP02(2019)139}{\emph{JHEP} {\bfseries 02}
  (2019) 139} [\href{https://arxiv.org/abs/1810.03818}{{\ttfamily
  1810.03818}}].

\bibitem{Frellesvig:2019uqt}
H.~Frellesvig, F.~Gasparotto, M.K.~Mandal, P.~Mastrolia, L.~Mattiazzi and
  S.~Mizera, \emph{{Vector Space of Feynman Integrals and Multivariate
  Intersection Numbers}},
  \href{https://doi.org/10.1103/PhysRevLett.123.201602}{\emph{Phys. Rev. Lett.}
  {\bfseries 123} (2019) 201602}
  [\href{https://arxiv.org/abs/1907.02000}{{\ttfamily 1907.02000}}].

\bibitem{evolutionary_programming}
A.~Eiben and J.~Smith, \emph{Introduction to Evolutionary Computing}, Springer
  Berlin, Heidelberg (2015),
  \href{https://doi.org/10.1007/978-3-662-44874-8}{10.1007/978-3-662-44874-8}.

\bibitem{genetic_programming}
J.R.~Koza, \emph{Genetic programming as a means for programming computers by
  natural selection},
  \href{https://doi.org/10.1007/BF00175355}{\emph{Statistics and Computing}
  {\bfseries 4} (1994) 87}.

\bibitem{funsearch}
B.~Romera-Paredes, M.~Barekatain, A.~Novikov, M.~Balog, M.P.~Kumar, E.~Dupont
  et~al., \emph{Mathematical discoveries from program search with large
  language models},
  \href{https://doi.org/10.1038/s41586-023-06924-6}{\emph{Nature} {\bfseries
  625} (2024) 468}.

\bibitem{deap}
F.-A.~Fortin, F.-M.~{De Rainville}, M.-A.~Gardner, M.~Parizeau and C.~Gagn\'e,
  \emph{{DEAP}: Evolutionary algorithms made easy}, {\emph{Journal of Machine
  Learning Research} {\bfseries 13} (2012) 2171}.

\bibitem{tHooft:1972tcz}
G.~'t~Hooft and M.J.G.~Veltman, \emph{{Regularization and Renormalization of
  Gauge Fields}},
  \href{https://doi.org/10.1016/0550-3213(72)90279-9}{\emph{Nucl. Phys. B}
  {\bfseries 44} (1972) 189}.

\bibitem{Abreu:2022mfk}
S.~Abreu, R.~Britto and C.~Duhr, \emph{{The SAGEX review on scattering
  amplitudes Chapter 3: Mathematical structures in Feynman integrals}},
  \href{https://doi.org/10.1088/1751-8121/ac87de}{\emph{J. Phys. A} {\bfseries
  55} (2022) 443004} [\href{https://arxiv.org/abs/2203.13014}{{\ttfamily
  2203.13014}}].

\bibitem{Lee:2008tj}
R.N.~Lee, \emph{{Group structure of the integration-by-part identities and its
  application to the reduction of multiloop integrals}},
  \href{https://doi.org/10.1088/1126-6708/2008/07/031}{\emph{JHEP} {\bfseries
  07} (2008) 031} [\href{https://arxiv.org/abs/0804.3008}{{\ttfamily
  0804.3008}}].

\bibitem{Kosower:2018obg}
D.A.~Kosower, \emph{{Direct Solution of Integration-by-Parts Systems}},
  \href{https://doi.org/10.1103/PhysRevD.98.025008}{\emph{Phys. Rev. D}
  {\bfseries 98} (2018) 025008}
  [\href{https://arxiv.org/abs/1804.00131}{{\ttfamily 1804.00131}}].

\bibitem{Maierhofer:2018gpa}
P.~Maierh\"ofer and J.~Usovitsch, \emph{{Kira 1.2 Release Notes}},
  \href{https://arxiv.org/abs/1812.01491}{{\ttfamily 1812.01491}}.

\bibitem{vonManteuffel:2014ixa}
A.~von Manteuffel and R.M.~Schabinger, \emph{{A novel approach to integration
  by parts reduction}},
  \href{https://doi.org/10.1016/j.physletb.2015.03.029}{\emph{Phys. Lett. B}
  {\bfseries 744} (2015) 101}
  [\href{https://arxiv.org/abs/1406.4513}{{\ttfamily 1406.4513}}].

\bibitem{Peraro:2016wsq}
T.~Peraro, \emph{{Scattering amplitudes over finite fields and multivariate
  functional reconstruction}},
  \href{https://doi.org/10.1007/JHEP12(2016)030}{\emph{JHEP} {\bfseries 12}
  (2016) 030} [\href{https://arxiv.org/abs/1608.01902}{{\ttfamily
  1608.01902}}].

\bibitem{Klappert:2019emp}
J.~Klappert and F.~Lange, \emph{{Reconstructing rational functions with
  FireFly}}, \href{https://doi.org/10.1016/j.cpc.2019.106951}{\emph{Comput.
  Phys. Commun.} {\bfseries 247} (2020) 106951}
  [\href{https://arxiv.org/abs/1904.00009}{{\ttfamily 1904.00009}}].

\bibitem{10.5555/645513.657604}
M.d.l.~Maza and B.~Tidor, \emph{An analysis of selection procedures with
  particular attention paid to proportional and boltzmann selection},  in
  \emph{Proceedings of the 5th International Conference on Genetic Algorithms},
  (San Francisco, CA, USA), p.~124–131, Morgan Kaufmann Publishers Inc.,
  1993.

\bibitem{copilot}
\url{https://news.microsoft.com/september-2023-event/}.

\bibitem{codey}
\url{https://lablab.ai/tech/google/codey}.

\bibitem{codellama}
B.~Rozière, J.~Gehring, F.~Gloeckle, S.~Sootla, I.~Gat, X.E.~Tan et~al.,
  \emph{Code llama: Open foundation models for code},
  \href{https://arxiv.org/abs/2308.12950}{{\ttfamily 2308.12950}}.

\bibitem{codestral}
\url{https://mistral.ai/en/news/codestral}.

\bibitem{funsearchfork}
J.~Aalto. Available at \url{https://github.com/jonppe/funsearch}.

\bibitem{podman}
``Podman: A tool for managing oci containers and pods.'' Available at
  \url{https://github.com/containers/podman}.

\bibitem{docker}
D.~Merkel, \emph{Docker: lightweight linux containers for consistent
  development and deployment}, {\emph{Linux J.} {\bfseries 2014} (2014) }.

\bibitem{llm}
S.~Willison, ``llm: Access large language models from the command-line.''
  Available at \url{https://github.com/simonw/llm}.

\bibitem{apptainer}
``Apptainer: Application containers for linux.'' Available at
  \url{https://github.com/apptainer/apptainer}.

\bibitem{singularity}
G.M.~Kurtzer, V.~Sochat and M.W.~Bauer, \emph{Singularity: Scientific
  containers for mobility of compute},
  \href{https://doi.org/10.1371/journal.pone.0177459}{\emph{PLOS ONE}
  {\bfseries 12} (2017) 1}.

\bibitem{Dersy:2022bym}
A.~Dersy, M.D.~Schwartz and X.~Zhang, \emph{{Simplifying Polylogarithms with
  Machine Learning}},
  \href{https://doi.org/10.1142/S2810939223500028}{\emph{Int. J. Data Sci.
  Math. Sci.} {\bfseries 1} (2024) 135}
  [\href{https://arxiv.org/abs/2206.04115}{{\ttfamily 2206.04115}}].

\bibitem{Cai:2024znx}
T.~Cai, G.W.~Merz, F.~Charton, N.~Nolte, M.~Wilhelm, K.~Cranmer et~al.,
  \emph{{Transforming the bootstrap: using transformers to compute scattering
  amplitudes in planar $\mathcal{N} = 4$ super Yang\textendash{}Mills theory}},
  \href{https://doi.org/10.1088/2632-2153/ad743e}{\emph{Mach. Learn. Sci.
  Tech.} {\bfseries 5} (2024) 035073}
  [\href{https://arxiv.org/abs/2405.06107}{{\ttfamily 2405.06107}}].

\bibitem{Cheung:2024svk}
C.~Cheung, A.~Dersy and M.D.~Schwartz, \emph{{Learning the Simplicity of
  Scattering Amplitudes}},  \href{https://arxiv.org/abs/2408.04720}{{\ttfamily
  2408.04720}}.

\bibitem{Koay:2025bmu}
Y.S.~Koay, R.~Enberg, S.~Moretti and E.~Camargo-Molina, \emph{{Generating
  particle physics Lagrangians with transformers}},
  \href{https://arxiv.org/abs/2501.09729}{{\ttfamily 2501.09729}}.

\bibitem{in_progress}
J.~Berman, F.~Charton, M.~von Hippel and M.~Wilhelm. In progress.

\end{thebibliography}\endgroup
\bibliographystyle{JHEP}

\end{document}